\documentclass[a4paper,11pt]{article}%
\usepackage{geometry}
\usepackage{amsmath}
\usepackage{amsfonts}
\usepackage{amssymb}
\usepackage{graphicx}
\usepackage{indentfirst}
\usepackage{bbold}
\usepackage[small,bf]{caption}
\usepackage{slashed}
\providecommand{\U}[1]{\protect\rule{.1in}{.1in}}

\begin{document}

\date{}
\title{\textbf{On the irrelevance of the Gribov issue in  $\mathcal{N}=4$ Super Yang--Mills in the Landau gauge}}
\author{\textbf{M.~A.~L.~Capri}\thanks{caprimarcio@gmail.com}\,\,,
\textbf{M.~S.~Guimaraes}\thanks{msguimaraes@uerj.br}\,\,,
\textbf{I.~F.~Justo }\thanks{igorfjusto@gmail.com}\,\,,\\
\textbf{L.~F.~Palhares}\thanks{leticiapalhares@gmail.com}\,\,,
\textbf{S.~P.~Sorella}\thanks{silvio.sorella@gmail.com}\,\,\,,\\[2mm]
{\small \textnormal{  \it Departamento de F\'{\i }sica Te\'{o}rica, Instituto de F\'{\i }sica, UERJ - Universidade do Estado do Rio de Janeiro,}}
 \\ \small \textnormal{ \it Rua S\~{a}o Francisco Xavier 524, 20550-013 Maracan\~{a}, Rio de Janeiro, Brasil}\normalsize}
\maketitle

\begin{abstract}
The issue of the Gribov copies is discussed in Euclidean $\mathcal{N}=4$ Super Yang-Mills theory quantized in the Landau gauge. As a consequence of the absence of a renormalization group invariant scale, it turns out that it is not possible to attach a dynamical meaning to the Gribov parameter. This implies that, in $\mathcal{N}=4$ Super Yang-Mills, there is no need for the restriction of the domain of integration in the functional integral to the Gribov region, and no non-perturbative dynamical scale is generated. This result is in agreement with the absence of a confining phase observed from the Coulomb behaviour of the Wilson loop evaluated through the AdS/CFT correspondence. The non-renormalization theorem of the gluon-ghost-antighost vertex of the Landau gauge is also explicitly checked till three loops. 

\end{abstract}

\section{Introduction}

Since the seminal work by Maldacena \cite{Maldacena:1997}, $\mathcal{N}=4$ Super Yang-Mills has become object of  intense investigation. The  AdS/CFT correspondence enables one to investigate the behaviour of the theory at large 't Hooft coupling by performing week coupling calculations in the corresponding dual supergravity theory in anti-de Sitter space. A central ingredient of this correspondence is the conformal character of $\mathcal{N}=4$ Super Yang-Mills, which exhibits vanishing $\beta$-function to all orders \cite{Ferrara:1974pu,Jones:1977zr,Grisaru:1979wc,Grisaru:1980nk,Caswell:1980ru,Velizhanin:2010vw}. This correspondence has allowed for the evaluation of correlation functions of gauge invariant operators at large 't Hooft coupling, probing thus in a simple and powerful way the non-perturbative region of a non-abelian gauge theory. This is the case of the Wilson loop, which displays a Coulomb behaviour \cite{Maldacena:1998im}, implying the absence of a confining phase, as dictated by the conformal character of the theory. \\\\On the other hand, as any renormalizable gauge theory, $\mathcal{N}=4$ Super Yang-Mills requires the introduction of the gauge-fixing and of the corresponding Faddeev-Popov term in order to be quantized. The theory is thus, in principle, affected by the presence of the Gribov copies \cite{Gribov:1977wm}. This is a non-perturbative effect, whose existence is related to the impossibility of picking up only one gauge configuration for each gauge orbit through a local and covariant  gauge-fixing procedure \cite{Singer:1978dk}. The Gribov issue has been much investigated in recent years in order to unravel aspects of confinement in Yang-Mills theories\footnote{See refs.\cite{Sobreiro:2005ec,Vandersickel:2012tz} for a pedagogical introduction to the Gribov problem.}. Nowadays, the so-called Gribov-Zwanziger framework \cite{Gribov:1977wm,Zwanziger:1988jt,Zwanziger:1989mf,Zwanziger:1992qr} enables us to take into account the existence of the Gribov copies within a local and renormalizable field theory framework. More precisely, it turns out that, in the Landau gauge,  the existence of the Gribov copies can be taken into account by restricting the domain of integration in the Euclidean functional integral to the Gribov region, as suggested by Gribov in his original work \cite{Gribov:1977wm}. Remarkably, this restriction results in a local and renormalizable quantum field theory framework, encoded in the so-called Gribov-Zwanziger action. The Gribov-Zwaniger construction is based on well established properties of the Gribov region and can be applied to any non-abelian gauge theory quantized in the Landau gauge, see \cite{Capri:2014xea} for a recent analysis of this issue in $\mathcal{N}=1$ Super Yang-Mills. The physical result of the restriction to the Gribov region $\Omega$ is the appearance of a mass parameter $\gamma$, known as the Gribov parameter. This parameter is not a free parameter of the theory. It has a dynamical meaning, being determined in a self-consistent way through a gap-equation which allows to express $\gamma$  in terms of the coupling constant and of the renormalization group invariant scale of the theory. The presence of this parameter deeply modifies the correlation functions of the theory in the non-perturbative infrared region and has a deep connection with the existence of a confining phase: the resulting gauge-boson two-point function violates the Osterwalder-Schrader axiom of reflection positivity \cite{Osterwalder:1973dx}, being thus incompatible with the construction of a standard K\"all\'en-Lehmann representation and with the propagation of a physical particle, as expected for a confined degree of freedom.
 \\\\It seems thus appropriate to investigate this issue in $\mathcal{N}=4$ Super Yang-Mills theory, in order to achieve  a different understanding of the absence of the confining phase. In particular, we shall be able to show that, due to the absence of a genuine renormalization group invariant scale, it is not possible to give a dynamical consistent meaning to the Gribov parameter. More precisely, it turns out that the only solution to the gap equation consistent with the renormalization group invariance of the vacuum energy is that corresponding to a vanishing Gribov parameter, {\it i.e.} $\gamma=0$.  This implies that, in the present case, there is no need to implement the restriction to the Gribov region $\Omega$,  in agreement with the absence of a confining phase as observed through the AdS/CFT correspondence.

\section{Euclidean $\mathcal{N}=4$ Super Yang-Mills theory and its restriction to the Gribov region} 

In $d=4$ Euclidean space-time the action of $\mathcal{N}=4$ Super Yang-Mills theory quantized in the Landau gauge can be written as 
\begin{equation}
S^{N=4}_{YM} = \frac{1}{4} \int d^4x \;F^a_{\mu\nu} F^a_{\mu\nu}   + \int d^4x \;{\cal{L}} (\lambda, \phi) + S_{gf} \;, \label{n4act}
\end{equation}
where, following \cite{Belitsky:2000ii}, ${\cal{L}} (\lambda, \phi)$ is given by 
\begin{eqnarray} 
{\cal{L}} (\lambda, \phi)  = & - & i {\bar \lambda}^{a \alpha'}_A ({\bar \sigma}_\mu)_{\alpha' \beta} D^{ab}_\mu \lambda^{b\beta A} + \frac{1}{4} (D_\mu^{ab} {\bar \phi}^b_{AB} ) (D_\mu^{ac} { \phi}^{cAB} )  - \frac{g}{\sqrt{2} }\left( f^{abc} {\bar \phi}^a_{AB} \lambda^{b\alpha A} \lambda^{cB}_{\alpha}  +   f^{abc} { \phi}^a_{AB} {\bar \lambda}^{b\alpha A} {\bar \lambda}^{cB}_{\alpha}   \right)  \nonumber \\
&+& \frac{g^2}{16} f^{abc} f^{amn} { \phi}^{bAB} { \phi}^{cCD}  {\bar \phi}^m_{AB} {\bar \phi}^n_{CD}  \;.   \label{n4}
\end{eqnarray}
In expression \eqref{n4}, the color indices, $a,b,c = 1,...,N^2-1$, correspond to the gauge group $SU(N)$. The indices $\alpha, \alpha'=1,2$ are spinor indices, while $A,B,C,D$ refer to the internal symmetry group $SO(5,1)$ obtained through the Wick rotation of $SO(6)$ from Minkowski to Euclidean space-time. The scalar field ${ \phi}^{bAB}$ is anti-symmetric in the indices $A,B$ with 
\begin{equation} 
{\bar \phi}^{a}_{AB} = \frac{1}{2} \epsilon_{ABCD}\;\phi^{aCD}  \;. \label{phi}
\end{equation} 
In the present analysis we shall consider the theory in the so-called super-conformal branch, characterized by a vanishing {\it vev} of the scalar field, $\langle \phi^a_{AB} \rangle=0$. \\\\As discussed in \cite{Belitsky:2000ii}, the action \eqref{n4act} can be obtained through dimensional reduction of  ten dimensional $\mathcal{N}=1$ super-Yang-Mills theory.  To some extent, expression \eqref{n4} can be regarded as a kind of matter Lagrangian with a particular content of the matter fields $(\lambda, \phi)$, all in the adjoint representation of the gauge group. \\\\The term $S_{gf}$ in eq.\eqref{n4act} stands for the Landau gauge-fixing term, namely 
\begin{equation}
S_{gf} = \int d^{4}x \left(  b^{a}\partial_{\mu}A^{a}_{\mu}
+\bar{c}^{a} \partial_{\mu}D^{ab}_{\mu}c^{b}  \right) \;,  \label{gf}
\end{equation}
where $({\bar c}^a, c^a)$ are the Faddeev-Popov ghosts, $b^a$ is the Lagrange multiplier implementing the Landau gauge, $\partial_\mu A^a_\mu=0$, and $D^{ab}_\mu =( \delta^{ab}\partial_\mu + g f^{acb}A^{c}_{\mu})$ is the covariant derivative in the adjoint representation of $SU(N)$.  \\\\Besides being renormalizable\footnote{An all order proof of the renormalizability of $\mathcal{N}=4$ Super Yang-Mills can be given by means of the use of an extended BRST operator which collects together both gauge and supersymmetry transformations, see \cite{White:1992wu,Maggiore:1994dw,Capri:2014jqa,Brandt:2002pa}  for applicantions to $\mathcal{N}=4,2,1$ Super Yang-Mills theories. }, the action \eqref{n4act} displays vanishing beta function \cite{Ferrara:1974pu,Jones:1977zr,Grisaru:1979wc,Grisaru:1980nk,Caswell:1980ru,Velizhanin:2010vw}, $\beta=0$, to all orders\footnote{See refs.\cite{White:1992wu,Baulieu:2006ru} for an all order proof of the vanishing of the $\beta$ function by using algebraic renormalization.}, {\it i.e.} expression \eqref{n4act} defines a conformal field theory, a fact which is at the basis of the AdS/CFT correspondence. \\\\Though, as any other gauge theory, the action $S^{N=4}_{YM}$ is plagued by the  Gribov copies \cite{Gribov:1977wm}, whose existence is a general feature of the gauge-fixing procedure \cite{Singer:1978dk}.  As already mentioned, in the Landau gauge, the issue of the Gribov copies can be handled by means of the Gribov-Zwanziger framework \cite{Gribov:1977wm,Zwanziger:1988jt,Zwanziger:1989mf,Zwanziger:1992qr}, which amounts to restrict  the domain of integration in the Euclidean functional integral to the so-called Gribov region $\Omega$, which is defined as the set of all gauge  
field configurations fulfilling the Landau gauge, $\partial_\mu A^{a}_\mu=0$,  and for which the Faddeev-Popov operator ${\cal M}^{ab}=-(\partial^2 \delta^{ab} -g f^{abc}A^{c}_{\mu}\partial_{\mu})$ is strictly positive, namely 
\begin{align}
\Omega \;= \; \{ A^a_{\mu}\;; \;\; \partial_\mu A^a_{\mu}=0\;; \;\; {\cal M}^{ab}=-(\partial^2 \delta^{ab} -g f^{abc}A^{c}_{\mu}\partial_{\mu})\; >0 \; \} \;. \label{gr}
\end{align} 
Following \cite{Gribov:1977wm,Zwanziger:1988jt,Zwanziger:1989mf,Zwanziger:1992qr}, the restriction of the domain of integration in the path integral is achieved by adding to the starting action an additional term $H(A)$, called the horizon term, given by the following non-local expression 
\begin{align}
H(A)  =  {g^{2}}\int d^{4}x\;d^{4}y\; f^{abc}A_{\mu}^{b}(x)\left[ {\cal M}^{-1}\right]^{ad} (x,y)f^{dec}A_{\mu}^{e}(y)   \;,  \label{hf1}
\end{align}
where ${\cal M}^{-1}$ stands for the inverse of the Faddeev-Popov operator. \\\\For the partition function of the theory implementing the restriction to the region $\Omega$ one thus writes 
\begin{equation}
 Z= \;
\int_\Omega {\cal D}A\; {\cal D}{\lambda} \; {\cal D}\phi\;{\cal D}c\;{\cal D}\bar{c}\; {\cal D} b\; e^{-S^{N=4}_{YM}}  =   \int {\cal D}A\;{\cal D}{\lambda} \; {\cal D}\phi\;{\cal D}c\;{\cal D}\bar{c} \; {\cal D} b \; e^{-(S^{N=4}_{YM}+\gamma^4 H(A) -V\gamma^4 4(N^2-1))} 
\;, \label{zww1}
\end{equation}
where $V$ is the Euclidean space-time volume. The parameter $\gamma$ has the dimension of a mass and is known as the Gribov parameter. It is not a free parameter of the theory. It is a dynamical quantity, being determined in a self-consistent way through a gap equation called the horizon condition \cite{Gribov:1977wm,Zwanziger:1988jt,Zwanziger:1989mf,Zwanziger:1992qr}, given by 
\begin{equation}
\left\langle H(A)   \right\rangle_{\Omega} = 4V \left(  N^{2}-1\right) \;,   \label{hc1}
\end{equation}
where the notation $\left\langle H(A)  \right\rangle_{\Omega}$ means that the vacuum expectation value of the horizon function $H(A)$ has to be evaluated with the measure defined in eq.\eqref{zww1}. It is worth underlining here that the expression of the horizon function $H(A)$, eq.\eqref{hf1}, enjoys a universal character, being independent from the kind of matter Lagrangian ${\cal{L}} (\lambda, \phi) $ which is added to the Yang-Mills term $\frac{1}{4} \int d^4x \;F^a_{\mu\nu} F^a_{\mu\nu} $. This important property follows from the fact that expression \eqref{hf1} can be derived through the Gribov no-pole condition \cite{Gribov:1977wm}, which amounts to evaluate the ghost two-point function in an external gauge field, a calculation which can be worked out in an exact fashion to all orders \cite{Capri:2012wx}. Since the Faddeev-Popov ghosts $({\bar c}, c)$ do not interact directly with the fields $(\lambda, \phi)$, {\it i.e.} interaction terms of the type $\sim$  $({\bar c}c \lambda, {\bar c} c\phi)$ are absent in expression \eqref{n4act},  and since the gauge field is considered as an external non-propagating field in the evaluation of the horizon function  \eqref{hf1}, it turns out that  $H(A)$ is independent from the matter terms. Of course, the matter fields will contribute to the gap-equation \eqref{hc1}, as in the evaluation of the vacuum expectation value $\left\langle H(A)   \right\rangle_{\Omega} $ the gauge field $A^a_\mu$ is no more considered as an external field, being in fact a fully propagating interacting field.  \\\\Although the horizon term  $H(A)$, eq.\eqref{hf1}, is non-local, it can be cast in local form by means of the introduction of a set of auxiliary fields $(\bar{\omega}_\mu^{ab}, \omega_\mu^{ab}, \bar{\varphi}_\mu^{ab},\varphi_\mu^{ab})$, where $(\bar{\varphi}_\mu^{ab},\varphi_\mu^{ab})$ are a pair of bosonic fields, while $(\bar{\omega}_\mu^{ab}, \omega_\mu^{ab})$ are anti-commuiting. It is not difficult to show that the partition function $Z$  in eq.\eqref{zww1} can be rewritten as \cite{Zwanziger:1988jt,Zwanziger:1989mf,Zwanziger:1992qr}
\begin{equation}
 Z = \;
\int {\cal D}A\;{\cal D}{\lambda} \; {\cal D}\phi\; {\cal D}c\;{\cal D}\bar{c}\; {\cal D} b \; {\cal D}{\bar \omega}\; {\cal D} \omega\; {\cal D} {\bar \varphi} \;{\cal D} \varphi \; e^{-S_{loc}} \;, \label{lzww1}
\end{equation}
where $S_{loc}$ is given by the local expression 
\begin{equation} 
S_{loc} = S^{N=4}_{YM}  + S_0+S_\gamma  \;, \label{sgz}
\end{equation}
with
\begin{equation}
S_0 =\int d^{4}x \left( {\bar \varphi}^{ac}_{\mu} (-\partial_\nu D^{ab}_{\nu} ) \varphi^{bc}_{\mu} - {\bar \omega}^{ac}_{\mu}  (-\partial_\nu D^{ab}_{\nu} ) \omega^{bc}_{\mu}  + gf^{amb} (\partial_\nu  {\bar \omega}^{ac}_{\mu} ) (D^{mp}_{\nu}c^p) \varphi^{bc}_{\mu}  \right) \;, \label{s0}
\end{equation}
and 
\begin{equation}
S_\gamma =\; \gamma^{2} \int d^{4}x \left( gf^{abc}A^{a}_{\mu}(\varphi^{bc}_{\mu} + {\bar \varphi}^{bc}_{\mu})\right)-4 \gamma^4V (N^2-1)\;. \label{hfl}
\end{equation} 
In the local formulation of the Gribov-Zwanziger action, the horizon condition \eqref{hc1} takes the simpler form 
\begin{equation}
 \frac{\partial \mathcal{E}_v}{\partial\gamma^2}=0\;,   \label{ggap}
\end{equation}
where $\mathcal{E}_{v}(\gamma)$ is the vacuum energy defined by:
\begin{equation}
 e^{-V\mathcal{E}_{v}}=\;Z\;  \label{vce} \;.
\end{equation}
In addition of being renormalizable to all orders, the action $S_{loc}$, eq.\eqref{sgz}, enjoys a certain number of non-renormalization theorems, namely 
\begin{eqnarray} 
Z_g\; Z_A^{1/2} Z_c  =   1 \;, \label{n1}  \\[3mm]
Z_{\gamma^2}  =   Z_g^{-1/2} Z_A^{-1/4} \;, \label{n2} 
\end{eqnarray}
where $(Z_g, Z_{\gamma^2}, Z_A, Z_c)$ stand, respectively, for the renormalization factors of the gauge coupling $g$, of the Gribov parameter $\gamma^2$, and of the gauge and ghost fields\footnote{We employ here the usual conventions for the renormalization factors and anomalous dimensions. For a generic field $\varphi$, we have $\varphi_0 = Z^{1/2}_\varphi \varphi$. The corresponding anomalous dimension is $\gamma_\varphi = {\bar \mu} \partial_{\bar \mu}\log Z^{1/2}_\varphi  $, where ${\bar \mu}$ is the renormalization scale.  For the gauge coupling constant $g$ we have: $g_0=Z_g g$. For the $\beta$-function: $\beta(g^2) = {\bar \mu} \partial_{\bar \mu}g^2 = -2 g^2 {\bar \mu} \partial_{\bar \mu} \log Z_g$. Finally, for the Gribov parameter $\gamma^2$: $\gamma^2_0 = Z_{\gamma^2} \gamma^2$ and  $ \gamma_{\gamma^2} = - {\bar \mu} \partial_{\bar \mu} \log Z_{\gamma^2}$.}. \\\\Equation \eqref{n1} states the  non-renormalization theorem of the gluon-ghost-antighost vertex, whose origin relies on the transversality of the gluon propagator in the Landau gauge. An all order proof of \eqref{n1} can be given within the algebraic renornalization set up  \cite{Piguet:1995er} thanks to the so-called ghost Ward identity, which is a general feature of the Landau gauge. The second equation \eqref{n2} expresses the fact that the renormalization factor $Z_{\gamma^2}$ of the Gribov parameter $\gamma^2$ is not an independent quantity of the theory. Again, property \eqref{n2}  can be proven to all orders thanks to the existence of a large set of Ward identities which can be established within the local formulation of the Zwanziger horizon function, see  \cite{Zwanziger:1988jt,Zwanziger:1989mf,Zwanziger:1992qr,Dudal:2008sp,Dudal:2011gd,Baulieu:2009xr} for algebraic proofs of \eqref{n2}. \\\\In the present case, due to the vanishing of the $\beta$-function,  {\it i.e.} $Z_g=1$, equations \eqref{n1},\eqref{n2} take the form
\begin{eqnarray} 
Z_A^{1/2} Z_c  =   1 \;, \label{nn1}  \\[3mm]
Z_{\gamma^2}  =    Z_A^{-1/4} = Z_c^{1/2}  \;. \label{nn2} 
\end{eqnarray}
Let us report here the three loop explicit expression of $Z_A, Z_c$, which can be found in the appendix of ref. \cite{Velizhanin:2008rw}, namely 
\begin{equation} 
Z_c =  1 + \frac{3}{4\epsilon} C_A a + \left( \frac{9}{32 \epsilon^2} - \frac{21}{32\epsilon} \right) C_A^2  a^2  +\left( \frac{9}{128 \epsilon^3} - \frac{189}{384\epsilon^2} + \frac{175}{192\epsilon} + \frac{79}{32 \epsilon} \zeta_3    \right) C_A^3  a^3  + O(a^4) \;, \label{ghn4}
\end{equation}
\begin{equation} 
Z_A = 1 - \frac{3}{2\epsilon} C_A a + \left( \frac{9}{8 \epsilon^2} + \frac{21}{16\epsilon} \right) C_A^2  a^2  +\left( -\frac{9}{16 \epsilon^3} - \frac{189}{96\epsilon^2} - \frac{175}{96\epsilon} - \frac{79}{16 \epsilon} \zeta_3    \right) C_A^3  a^3  + O(a^4)  \;, \label{gln4} 
\end{equation}
where $C_A=N$, $a=\frac{g^2}{16\pi^2}$ and $\epsilon$ stand for regularization parameter of the dimensional reduction scheme. From 
\begin{equation} 
\sqrt{1+x} = 1 +\frac{1}{2} x -\frac{1}{8}x^2 + \frac{1}{16}x^3 + O(x^4)   \;, \label{x} 
\end{equation} 
one  obtains
\begin{eqnarray} 
Z_A ^{1/2}&  = &  1 - \frac{3}{4\epsilon} C_A a + \left( \frac{9}{32 \epsilon^2} + \frac{21}{16\epsilon} \right) C_A^2  a^2    \nonumber \\
  \; \; \; \;  &+ & \left( -\frac{9}{32 \epsilon^3} - \frac{189}{192\epsilon^2} - \frac{175}{192\epsilon} - \frac{79}{32 \epsilon} \zeta_3  + \frac{27}{128 \epsilon^3} + \frac{63}{128 \epsilon^2} \right) C_A^3  a^3  + O(a^4)  \;, \label{ggln4} 
\end{eqnarray}
from which one immediately verifies that 
\begin{equation}
Z_A^{1/2} Z_c  =   1  + O(a^4) \;. \label{nnn1}
\end{equation}
Also, when translated at the level of the anomalous dimensions, equations \eqref{nn1},\eqref{nn2} take the form 
\begin{eqnarray} 
 \gamma_A(g^2)  & = &  - 2 \gamma_c(g^2) \;, \label{an1} \\[3mm] 
\gamma_{\gamma^2}(g^2)&  = &- \gamma_c(g^2) \;. \label{an2}
\end{eqnarray}

\subsection{Explicit verification of the renormalization group invariance of the vacuum energy at the first-order} 

Let us end this section by explicitly checking that, at the first order,  the vacuum energy $\mathcal{E}_{v}$ obtained from the local action $S_{loc}$, \eqref{sgz},  is  invariant under the renormalization group equations, namely 
\begin{equation}
\left( {\bar \mu} \frac{\partial}{\partial \bar \mu} + \gamma_{\gamma^2} \gamma^2 \frac{\partial}{\partial \gamma^2}  \right) \mathcal{E}_{v}=0 \;.   \label{regn4}
\end{equation}
As a consequence of the presence of the massive parameter $\gamma$, it turns out that the vacuum energy corresponding to the action $S_{loc}$ is non-vanishing. Using  dimensional regularization in the  $\overline{MS}$  scheme,  for the one-loop order vacuum energy $\mathcal{E}_{v}^{(1)}$,  one obtains \begin{equation} 
\mathcal{E}_{v}^{(1)} = - 4(N^2-1) \gamma^4 - \hbar \frac{3(N^2-1)}{64\pi^2} (2Ng^2 \gamma^4) \left(  \log \frac{2Ng^2\gamma^4}{{\bar \mu}^4} - \frac{8}{3} \right) \;, \label{vcn4}
\end{equation}
where we have introduced the factor $\hbar$ to make explicit the loop order of the various terms. From 
\begin{equation}
\gamma_{\gamma^2}^{(1)}(g^2) = - \gamma_c^{(1)}(g^2)= \frac{3N}{4} \frac{g^2}{16\pi^2} \;,  \label{ghan}
\end{equation}
it follows that
\begin{equation} 
\left( {\bar \mu} \frac{\partial}{\partial \bar \mu} + \hbar \gamma_{\gamma^2}^{(1)} \gamma^2 \frac{\partial}{\partial \gamma^2}  \right) \mathcal{E}_{v}^{(1)}= 
\hbar (N^2-1) \gamma^4 \left( \frac{24 N g^2}{64\pi^2} - \frac{24 N g^2}{64\pi^2} \right) + O(\hbar^2) = 0 + O(\hbar^2)  \;, \label{invrge} 
\end{equation}
which shows the renormalization group invariance of the vacuum energy at the first order. 
\section{Analysis of the vacuum energy and impossibility of attaching a dynamical meaning to the Gribov parameter $\gamma$} 
So far, the Gribov parameter $\gamma$ has been considered as a free parameter of the theory. However, we have already underlined that $\gamma$ is constrained by the horizon condition \eqref{ggap}, which enables us to express $\gamma$ as a function of the coupling constant $g$ and of the scale $\bar \mu$ in a non-perturbative way, {\it i.e.} $\gamma^2 \sim {\bar \mu}^2 e^{- \frac{const.}{g^2}}$. The horizon condition \eqref{ggap} plays a fundamental role in the Gribov-Zwanziger theory. It is precisely this condition which allows us to capture information on the non-perturbative sector of the theory. Though, due to the absence of a renormalization group invariant scale in $\mathcal{N}=4$ Super Yang-Mills, it turns out that the horizon condition \eqref{ggap}  is incompatible with the renormalization group invariance of the vacuum energy. In other words, the absence of an invariant scale makes  it impossible to attach a consistent dynamical meaning to the parameter $\gamma$. The only way out is that of imposing $\gamma=0$, which implies that the restriction of the integration domain  in the functional integral to the Gribov region is not needed in $\mathcal{N}=4$. This means that the Gribov issue is irrelevant in the case of $\mathcal{N}=4$ Super Yang-Mills. For the benefit of the reader, before discussing $\mathcal{N}=4$ Super Yang-Mills, let us proceed by showing how the whole framework works in the case of pure Yang-Mills theories, see for example  \cite{Dudal:2008sp}.   
\subsection{The case of pure Yang-Mills theory} 
In the case of pure Yang-Mills, the action \eqref{sgz} takes the form 
\begin{equation} 
S_{GZ} = \int d^4x \left( \frac{1}{4} F^a_{\mu\nu} F^a_{\mu\nu} + b^a \partial_\mu A^a_\mu + {\bar c}^a {\partial_\mu} D_{\mu}^{ab} c^b \right)  + S_0+S_\gamma  \;, \label{gz}
\end{equation}
where $S_0$ and $S_\gamma$ are given by expressions  \eqref{s0}, \eqref{hfl}. Expression \eqref{gz} is known as the Gribov-Zwanziger action \cite{Gribov:1977wm,Zwanziger:1988jt,Zwanziger:1989mf,Zwanziger:1992qr}. In the present case, the theory has a non-vanishing $\beta$-function, and the vacuum energy obeys the renormalization group equation 
\begin{equation}
\left( {\bar \mu} \frac{\partial}{\partial \bar \mu} + \gamma_{\gamma^2} \gamma^2 \frac{\partial}{\partial \gamma^2} + \beta(g^2)\frac{\partial}{\partial g^2}  \right) \mathcal{E}_{v}^{GZ}=0 \;.   \label{reggz}
\end{equation}
Moreover, due to the non-vanishing of the $\beta$-function, the theory exhibits a genuine renormalization invariant scale $\Lambda_{YM}$, defined through the equation 
\begin{equation}
\left( {\bar \mu} \frac{\partial}{\partial \bar \mu}  + \beta(g^2)\frac{\partial}{\partial g^2}  \right) \Lambda_{YM}=0 \;.   \label{lambda}
\end{equation}
At the first order, eq.\eqref{lambda} gives the well known result 
\begin{equation}
\Lambda_{YM} = {\bar \mu} \; e^{-\frac{3}{22} \frac{16\pi^2}{Ng^2} }   \;, \label{lambda1}
\end{equation}
 where use has been made of 
 \begin{equation} 
 \beta^{(1)}(g^2)= - \frac{22}{3} \frac{g^4 N}{16\pi^2} \;. \label{beta1}
 \end{equation}
 Let us now impose that the Gribov parameter $\gamma$ is determined by the horizon condition, namely 
 \begin{equation}
 \frac{\partial \mathcal{E}_v^{GZ}}{\partial\gamma^2}=0\;.   \label{gzgap}
\end{equation}
Therefore, one immediately realizes that, due to eq.\eqref{gzgap}, equation  \eqref{reggz} becomes 
 \begin{equation}
\left( {\bar \mu} \frac{\partial}{\partial \bar \mu} + \beta(g^2)\frac{\partial}{\partial g^2}  \right) \mathcal{E}_{v}^{GZ}=0 \;,   \label{rgz1}
\end{equation}
 which tells us that the vacuum energy is a renormalization group invariant quantity, namely $ \mathcal{E}_{v}^{GZ} \sim \Lambda^4_{YM}$, see \cite{Dudal:2008sp}.

\subsection{The case of N=4 Super Yang-Mills theory} 
Let us face now the case of $\mathcal{N}=4$ Super Yang-Mills. As a consequence of the vanishing of the $\beta$-function, the theory does not possess a renomalization group invariant scale, {\it i.e.} there is no analogue of $\Lambda_{YM}$, a fact which expresses in physical terms the conformal invariance of the theory.  One should therefore be able to consistently prove that the restriction to the Gribov region does not generate any new dynamical scale in this gauge theory.
\\\\With this aim, let us  try to repeat the procedure done in the previous case of Yang-Mills theory, by looking at the horizon condition \eqref{ggap}. From expression \eqref{vcn4}, we get  (at one loop)
\begin{equation} 
 \frac{\partial \mathcal{E}_v^{(1)}}{\partial\gamma^2}=0 \;\;\;\; \Rightarrow  \;\;\;\; \gamma^2 \left(  \left(2-\frac{5 Ng^2}{64\pi^2}\right) + \frac{3}{64\pi^2} Ng^2 \log \frac{2Ng^2\gamma^4}{{\bar \mu}^4} \right)  = 0 \;, 
\end{equation}
which gives the following two solutions 
\begin{eqnarray} 
\gamma^2 &=& 0 \;, \label{g1}  \\[3mm]
\gamma^4 & = & \frac{{\bar \mu}^4}{2Ng^2}  \; e^{ \frac{5}{3} } \; e^{-\frac{128\pi^2}{3Ng^2}}  \;. \label{g2} 
\end{eqnarray}
As discussed in \cite{Dudal:2008sp}, in pure Yang-Mills  the solution \eqref{g1} has to be rejected, and one keeps the second one, which gives a non-vanishing Gribov parameter.  However, in the present case, it is the second solution which has to be rejected since  it provides a solution for $\mathcal{E}_{v}$ that fails to satisfy the renormalization-group equation for the vacuum energy, eq.\eqref{regn4}.
Indeed, owing to eq.\eqref{regn4}, the horizon condition  \eqref{ggap} would imply that the vacuum energy $ \mathcal{E}_v$ is independent from the scale $\bar \mu$, namely 
\begin{equation}
 {\bar \mu} \frac{\partial}{\partial \bar \mu} \mathcal{E}_{v}=0 \;.   \label{i1}
\end{equation}
while, in the $\mathcal{N}=4$ Super Yang-Mills case, the second solution \eqref{g2}  would give a vacuum energy which manifestly depends on $\bar \mu$, {\it i.e.} 
\begin{equation} 
\mathcal{E}_v^{(1)} = \frac{3(N^2-1)}{64\pi^2} {\bar \mu}^4  \; e^{ \frac{5}{3} } \; e^{-\frac{128\pi^2}{3Ng^2}}  \;, \label{i2}
\end{equation}
being thus inconsistent with  \eqref{i1}. In the present case, we are left thus with the first solution, eq.\eqref{g1}, {\it i.e.} $\gamma^2=0$, which  gives a vanishing vacuum energy
\begin{equation}
\mathcal{E}_v^{(1)}= 0 \;, \label{z}
\end{equation}
in agreement with the conformal invariance of the theory. \\\\Setting $\gamma^2=0$  has the meaning of removing the restriction to the Gribov region $\Omega$ in the functional integral \eqref{lzww1}, as it follows by noticing that,  when $\gamma=0$, the integration over the auxiliary fields $(\bar{\omega}_\mu^{ab}, \omega_\mu^{ab}, \bar{\varphi}_\mu^{ab},\varphi_\mu^{ab})$ gives a unity. Therefore the existence of Gribov copies beyond the Gribov region in $\mathcal{N}=4$ Super Yang-Mills cannot be associated with the generation of a non-perturbative dynamical scale, as is the case of pure Yang-Mills, in agreement with the absence of a confining phase in the former theory.\\\\The same reasoning presented above can be in fact systematically extended to all orders, proving in an exact way the vanishing of the Gribov parameter and the fact that the Gribov issue in $\mathcal{N}=4$ Super Yang-Mills theories assumes, in contrast to the pure Yang-Mills case, an irrelevant role as far as the dynamical generation of a nonperturbative physical scale is concerned. The exact result follows from the fact that the renormalization group equation for the vacuum energy \eqref{regn4} in the presence of the horizon condition \eqref{ggap} assumes the form \eqref{i1} to all orders, which in turn is inconsistent with any nonvanishing expression for the vacuum energy depending on $\bar\mu$. Using the general form of the vacuum energy computed with the restriction to the Gribov region:
\begin{equation}
\mathcal{E}_v= \gamma^4 f\left(\frac{\gamma^2}{\bar\mu^2}\right) \;,
\end{equation}
one directly sees that the all-order Gribov gap equation yields:
\begin{equation} 
 \frac{\partial \mathcal{E}_v}{\partial\gamma^2}=0 \;\;\;\; \Rightarrow  \;\;\;\; \gamma^2 \left( 
 2 f\left(\frac{\gamma^2}{\bar\mu^2}\right)+\frac{\gamma^2}{\bar\mu^2} f'\left(\frac{\gamma^2}{\bar\mu^2}\right) \right)  = 0 \;, 
\end{equation}
implying that either the Gribov parameter and the vacuum energy vanish, or they must both be functions of $\bar\mu^2$. It is finally clear that, in this case, the only consistent exact result is $\gamma^2=0$ and $\mathcal{E}_v=0$.
\\\\Finally, it is worth to spend a few words on the ghost propagator $ {\cal G}^{ab}(p^2) = \langle {\bar c}^{a}(p) c^{b}(-p) \rangle  $. As it is apparent from eq.\eqref{gr},  the restriction to the Gribov region, defined by $\Omega$, implies that the  ghost propagator  $ {\cal G}^{ab}(p^2) $ is always positive. It cannot change sign as one varies the momentum $p^2$. A change of sign of $ {\cal G}^{ab}(p^2) $  would in fact imply that one has left the Gribov region $\Omega$, so that negative eigenvalues of the Faddeev-Popov operator ${\cal M}^{ab} $  show up. Nevertheless, in the present case we have shown that, as a consequence of the absence of an invariant physical scale,  there is no need to implement the restriction to the region $\Omega$. We should thus be able to consistently show that, in $N=4$ Super-yang-Mills, the ghost propagator  $ {\cal G}^{ab}(p^2) $ cannot undergo a change of sign as one varies $p^2$. This nice feature can be proven as a  consequence of the renormalization group equations of the theory in the Landau gauge. Parametrizing the ghost propagator in the standard way, {\it i.e.} 
\begin{equation} 
{\cal G}^{ab}(p^2)  = \frac{ \delta^{ab} }{p^2} \; {\cal G}\left( {\frac{{\bar \mu}^2} {p^2} }\right)   \;, \label{ghp}   
\end{equation}
it follows that, due to the vanishing of the $\beta$-function of the theory, the form factor ${\cal G}\left( {\frac{{\bar \mu}^2} {p^2} }\right) $ obeys the following Renormalization Group Equation 
\begin{equation}
\left( {\bar \mu} \frac{\partial}{\partial \bar \mu} - 2 \gamma_c  \right)  {\cal G} \left( {\frac{{\bar \mu}^2}{p^2} }\right) = 0 \;,    \label{ghp1}
\end{equation}
where $\gamma_c(g^2)$ is the ghost anomalous dimension. The vanishing of the $\beta$-function enables us to solve exactly this equation, giving 
\begin{equation}
{\cal G}\left( {\frac{{\bar \mu}^2} {p^2} }\right)  =  \eta(g^2)  {\left( {\frac{{\bar \mu}^2} {p^2} } \right)}^{\gamma_c(g^2)}  \;, \label{ghp2}
\end{equation}
where $\eta(g^2)$ is a dimensionless function of the coupling constant $g^2$. Even if we would not be able to compute exactly the ghost  anomalous dimension $\gamma_c(g^2)$ and the function $\eta(g^2)$,  eq.\eqref{ghp2} shows the important feature that the form factor ${\cal G}\left( {\frac{{\bar \mu}^2} {p^2} }\right) $ cannot change sign as one varies the momentum $p^2$. In particular, no extra poles at non-vanishing values of $p^2$ will show up, confirming that no restriction to the Gribov region $\Omega$ is in fact needed in  this theory.
\\\\In summary, in the case of $\mathcal{N}=4$ Super Yang-Mills theory, the vanishing of the beta function makes  it impossible to give a non-vanishing dynamical meaning to the Gribov parameter and the existence of Gribov copies cannot be associated with the generation of a non-perturbative dynamical  scale as in the case of pure Yang-Mills. \\\\Although in this work the theory has been considered in its super-conformal branch, the present framework can be applied as well to the study of the Gribov issue in the Coulomb phase, where the scalar field $\phi^a_{AB}$ displays a non-vanishing {\it vev}.  This analysis will be postponed for a future investigation. 

\section*{Acknowledgments}
The Conselho Nacional de Desenvolvimento Cient\'{\i}fico e
Tecnol\'{o}gico (CNPq-Brazil), the Faperj, Funda{\c{c}}{\~{a}}o de
Amparo {\`{a}} Pesquisa do Estado do Rio de Janeiro,  the
Coordena{\c{c}}{\~{a}}o de Aperfei{\c{c}}oamento de Pessoal de
N{\'{\i}}vel Superior (CAPES)  are gratefully acknowledged. L.F.P. is supported by a BJT fellowship from the brazilian program ``Ci\^encia sem Fronteiras''.

\end{document}